\documentclass[journal=ancac3,manuscript=article]{achemso}

\usepackage{amsmath,amssymb,amsfonts}
\usepackage{bm}
\usepackage{color, soul}
\usepackage{graphicx}
\usepackage{float}
\usepackage{wrapfig}
\usepackage{verbatim}
\usepackage[english]{babel}
\usepackage{hyperref}[colorlinks=true,bookmarks=false,citecolor=blue,urlcolor=blue]
\usepackage{natbib}
\usepackage{subfigure}
\usepackage{amstext, mathrsfs, textcomp}
\usepackage{multirow}
\usepackage{xcolor}
\usepackage{enumerate}
\usepackage{epstopdf}
\usepackage{tabularx}
\makeatletter

\usepackage{gensymb}

\newcommand*{\citen}[1]{%
  \begingroup
    \romannumeral-`\x 
    \setcitestyle{numbers}%
    \cite{#1}%
  \endgroup   
}

\setkeys{acs}{etalmode=truncate,maxauthors=0}
\graphicspath{{figs/}}

\author{Daniil Riabov}
\affiliation{Bionanophotonic Systems Laboratory, Institute of Bioengineering, School of Engineering, EPFL, Lausanne, Switzerland}
\email{daniil.riabov@epfl.ch}

\author{Abtin Saateh}
\affiliation{Bionanophotonic Systems Laboratory, Institute of Bioengineering, School of Engineering, EPFL, Lausanne, Switzerland}

\author{Wenhong Yang}
\affiliation{Bionanophotonic Systems Laboratory, Institute of Bioengineering, School of Engineering, EPFL, Lausanne, Switzerland}

\author{Ivan Sinev}
\affiliation{Bionanophotonic Systems Laboratory, Institute of Bioengineering, School of Engineering, EPFL, Lausanne, Switzerland}

\author{Yuri Kivshar}
\affiliation{Nonlinear Physics Center, Research School of Physics, Australian National University, Canberra ACT 2601, Australia}

\author{Hatice Altug}
\affiliation{Bionanophotonic Systems Laboratory, Institute of Bioengineering, School of Engineering, EPFL, Lausanne, Switzerland}
\email{hatice.altug@epfl.ch}

\title{
Digital nanophotonic biosensing empowered by silicon Mie voids
}

\begin{document}

\begin{abstract}

Optical biosensors are indispensable in medical and environmental diagnostics, yet existing approaches are fundamentally limited in their sensitivity due to ensemble-averaged measurements. Digital biosensing has emerged as a promising solution for resolving individual binding events, thereby providing signals at very low analyte concentrations down to the single-molecule level. Here, we present a novel concept for digital optical biosensing empowered by dielectric Mie voids, combining nanoparticle-based contrast enhancement and deep learning for ultrasensitive biomarker detection. The resonantly trapped light in the air cavities of the periodic Mie void arrays ensures strong overlap between the near-fields and the single gold nanoparticles that are captured on the surface in the presence of the protein biomarker. Remarkably, this strong interaction creates high-contrast digital signals for the precise counting of single nanoparticles located both within and outside the voids, yielding efficient use of the entire sensor area for high sensitivity. We employ deep-ultraviolet (DUV) lithography for the scalable and low-cost production of Mie voids in silicon wafers and automated image analysis with a convolutional neural network for robust nanoparticle counting. As a proof of our concept, we demonstrate the detection of an important disease biomarker, interleukin-6 (IL-6), from small sample volumes at concentrations as low as 1.84 pg/ml, within the physiological range of healthy individuals. Owing to its scalability, precision, and adaptability, our digital nanophotonic biosensing approach based on silicon Mie voids establishes a versatile route for applications ranging from bioanalytics to health and environmental monitoring.

\end{abstract}

\newpage

\section{Main}
Optical biosensors play a crucial role in both medical and environmental diagnostics, allowing the monitoring of minuscule concentrations of relevant biological entities from a variety of  samples~\cite{Chen2020, Uniyal2023}. Due to their robustness against external variations in temperature, conductivity, or humidity, sensing with optical readout methods has become ubiquitous, including fluorescence-based detection\cite{Mal2024}, infrared (IR) and Raman spectroscopy\cite{Adato2015, Serebrennikova2021}, surface plasmon resonance (SPR)\cite{Singh2016}, and waveguide and fiber sensing\cite{Butt2022}, among others. Driven by the demand for miniaturization and enhanced sensitivity, novel optical biosensing approaches leverage resonant nanophotonics\cite{monticone2017metamaterial,koshelev2020dielectric, baranov2017all}, which employ versatile metallic and dielectric platforms such as whispering-gallery-mode (WGM) resonators\cite{Jiang2020}, localized surface plasmon resonances (LSPR)\cite{Mayer2011}, photonic crystals\cite{Nair2010}, and high-$Q$ metasurfaces\cite{Zhang2021, Tseng2021}. The resonant nature of these systems facilitates the interaction of the locally enhanced electromagnetic field with an analyte, which manifests as an alteration of the ensemble optical characteristics, e.g. resonant wavelength shift, reflectance or transmission intensity change, etc. However, in this case, the performance of the platform is typically limited by the surface density and thickness of the adsorbed molecular biolayer that is required to induce a distinguishable perturbation. 

To mitigate this challenge, the concept of digital biosensing has emerged recently, in which each binding event is treated separately without ensemble averaging, thus allowing `to quantify specific target molecules with enough precision by counting individual events, and to observe the characteristics of such biomolecular interactions.''\cite{Huang2020}. Such an approach dramatically improves the sensitivity, ultimately leading down to single-molecule resolution. Digital optical sensing can be performed by employing single optical resonators \cite{Vollmer2008, Subramanian2020, Zijlstra2012, Beuwer2015}. While this approach allows tracking separate molecule binding events, it often suffers from low throughput measurements and susceptibility to external noise sources. These limitations can be addressed by implementing a contrast agent attached to the biomarker of interest. A common strategy involves using plasmonic nanoparticles (NPs) as such an agent. Analyte molecules bind to the surface of dispersed particles in the solution, and subsequently, these conjugates get immobilized on the sensor surface\cite{Jing2019, Belushkin2018, Belushkin2020, Canady2019, Che2019, VanDongen2022}. The distinct scattering and absorption properties of the nanoparticles enable a strongly localized perturbation in their vicinity, thus facilitating the counting of binding events. 

A general requirement for optical digital sensing platforms is a good spatial overlap between the target analyte and the field profile of the resonant mode. For this reason, many systems utilize plasmonic platforms that provide relatively accessible, strong local field enhancement at their surface. Dielectric materials, despite their potential for achieving ultra-high-$Q$ resonances\cite{Tian2020, Koshelev2018}, typically localize the field inside the material, therefore reducing the field interaction with the analyte in the surrounding medium and limiting the sensor's performance. Recently, a novel concept of Mie voids has emerged, where light is resonantly confined in air cavities within dielectric materials~\cite{Hentschel2023}. These voids are essentially localized nanogrooves supporting Mie-type resonances inside their volume. The field localization in air gives analytes direct access to the hotspots and allows for measurements of extremely small sample volumes\cite{Arslan2025} or for sizing and detecting micro- and nanoplastic particles\cite{ludescher2025optical}.

\begin{figure}[h!]
    \centering
    \includegraphics[width=1\columnwidth]{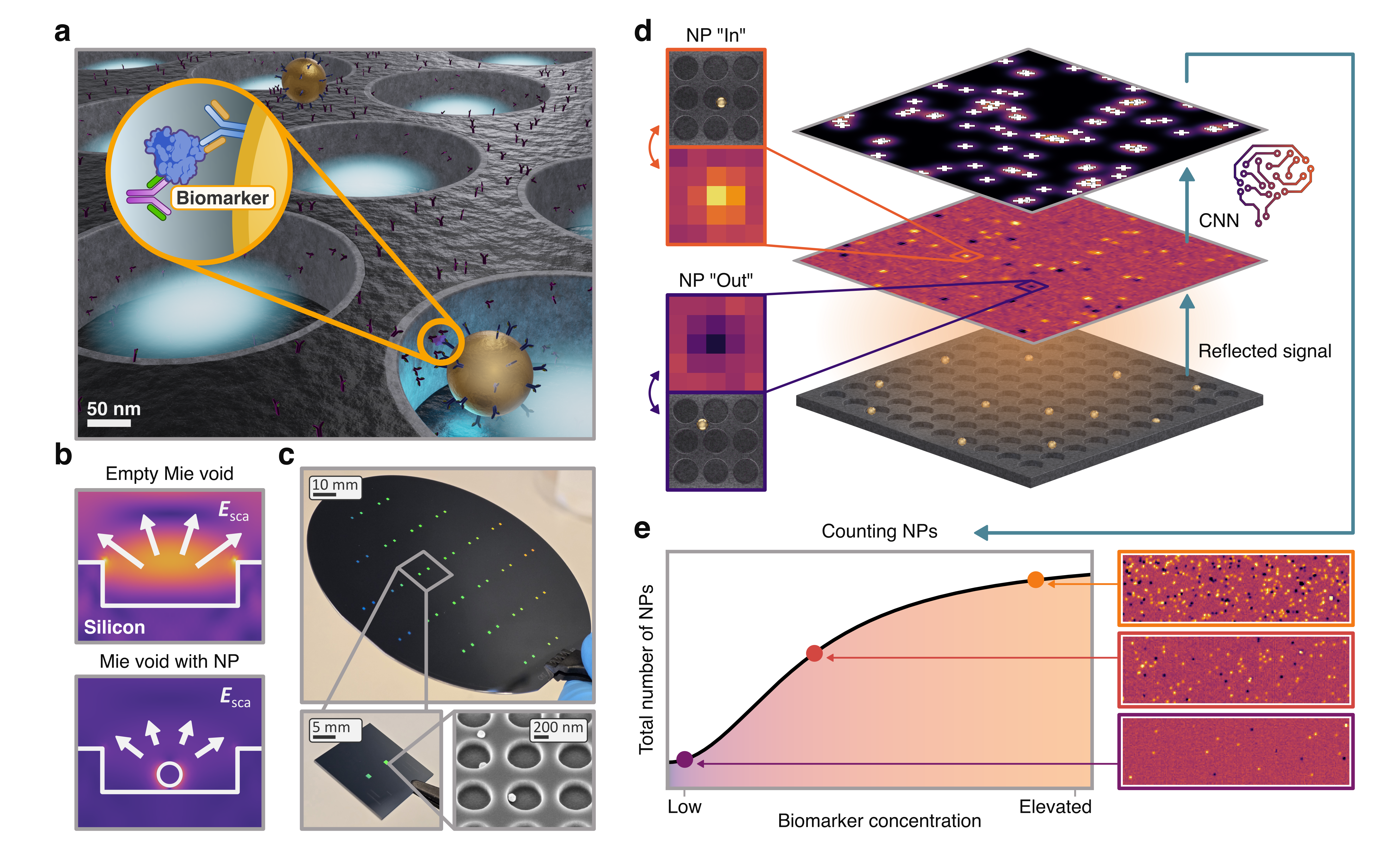}
    \caption{\textbf{CNN-enhanced digital biosensing with dielectric Mie voids.} a) Artistic view of a silicon Mie void sensor with bound gold nanoparticles (NPs). The biomarker molecule (blue) is 'sandwiched' between the detection (attached to the NP) and capture (attached to the sensor surface) antibodies, ensuring selective immobilization of the nanoparticle on the sensor. b) Simulation of the change in the scattering properties of a silicon Mie void upon binding of a gold NP. A decrease in the scattering cross section of the void leads to a detectable signal change in reflection for detecting and counting individual NPs. c) Top image: Photograph of a 4-inch silicon wafer containing 24 Mie void chips fabricated using DUV lithography; bottom left image: photograph of a single chip with Mie void arrays obtained by dicing the wafer; bottom right image: SEM image of Mie voids with bound gold NPs. d) Illustration of the CNN aided digital NP counting scheme for biomarker detection: incident light reflects from a Mie void array with Au NPs, forming an optical image on the CCD camera. Nanoparticles induce local perturbations in reflectivity, creating brighter (for those located inside ("In") the voids) or darker (for those located outside ("Out") of the voids) spots compared to the background. The resulting image is post-processed using a pre-trained CNN model to count the total number of binding events. e) Calibration curve showing the total number of detected NPs for different biomarker concentrations.}  
     \label{fig1:intro}
\end{figure}

In this article, we introduce an ultrasensitive digital biosensing approach using silicon Mie voids by counting individual nanoparticles that are immobilized on a high-index dielectric substrate through the binding of disease biomarkers (Fig.~\ref{fig1:intro}~(a)). The detection principle is based on imaging localized intensity changes induced by the strong interaction of resonantly trapped light in the voids with bound single gold nanoparticles near the surface (Fig.~\ref{fig1:intro}~(b)). Strikingly, the Mie void mode field extends to the regions around the voids, thus enabling detection of the particles located both inside and outside of the voids, which ensures efficient use of the sensor surface area for further improvement of sensitivity over some of the previously reported platforms. Immobilized nanoparticles generate brighter (NP "In") or darker (NP "Out") signals of high contrast on a homogeneously dark background resulting from the suppression of light reflection by the nanopatterned substrate. These signals are imaged over a wide field-of-view and processed using a convolutional neural network to automatically count numerous binding events with high precision and recall (Fig.~\ref{fig1:intro}~(d)). As a result, the presented approach provides the detection of $\sim$80\% of all the nanoparticles on the surface with as low as $\sim$10\% false positive instances. In addition, the critical dimensions of our design are compatible with CMOS fabrication methods, particularly, DUV lithography, which we employed for scalable and low-cost production of the sensor chips on 4-inch silicon wafers (Fig.~\ref{fig1:intro}~(c)). We experimentally showcase the application of the device for detecting ultra-low concentrations of interleukin-6 (IL-6), a key cancer and inflammation biomarker\cite{Mehta2024, Zhang2020, Lippitz2016}, in small sample volumes ($\sim5~\mu$l), by utilizing the surface functionalization strategies. Our sensor achieves a limit of detection (LoD) as low as 1.84 pg/ml, which corresponds to IL-6 concentrations in healthy individuals\cite{Said2021}. Elevated biomarker concentration leads to a larger number of bound nanoparticles, which is confirmed by our platform (Fig.~\ref{fig1:intro}~(e)). This novel nanophotonic digital biosensing platform, utilizing low-cost silicon chips and smart image analysis, is versatile and can be adapted to sensitively detect a variety of biomolecules for bioanalytical and diagnostics applications.

\section{Simulations and fabrication}
\begin{figure}[t!]
    \centering
    \includegraphics[width=1\columnwidth]{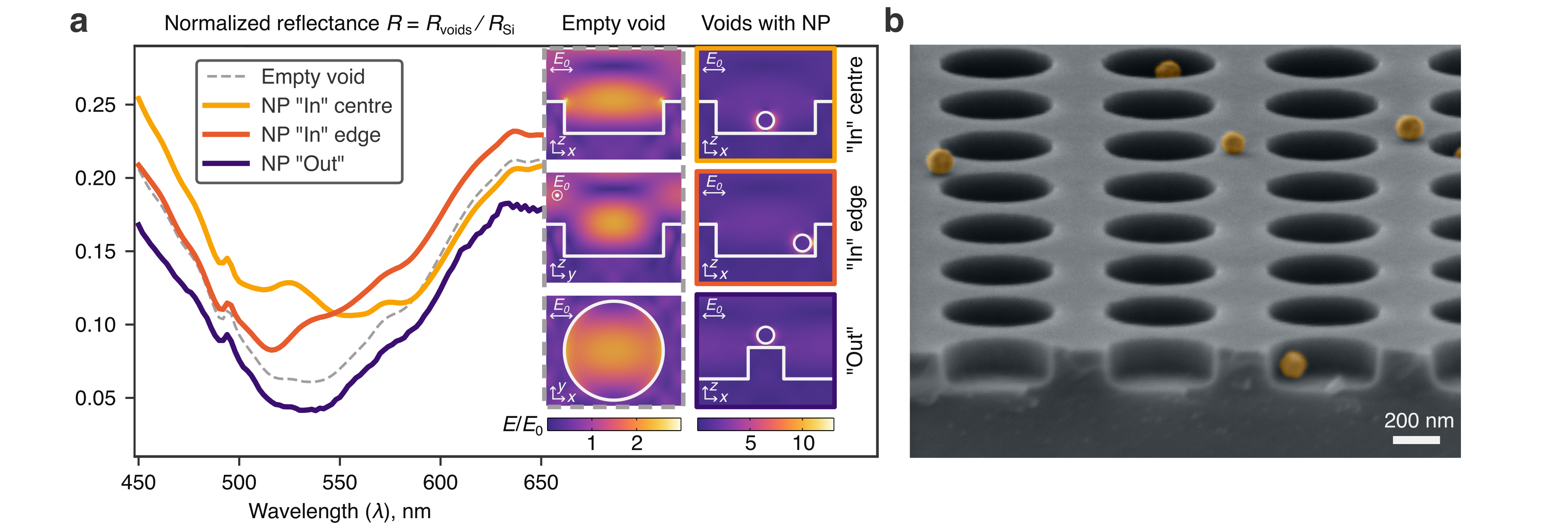}
    \caption{\textbf{Simulated reflectance spectra and fabricated structures.} a) Linearly polarized light reflectance spectra of the silicon Mie void array normalized to the reflectance from a bare silicon substrate. The reflectance signal increases when a gold NP is located inside ("In") of the void and depends on the NP position (at the centre or near the void edge). The reflectance signal decreases when a gold NP is outside ("Out") of the voids. The insets show the corresponding electric field distributions of the void, with the polarization direction shown by an arrow in each inset. b) False color SEM image of a DUV-fabricated Mie void array with bound gold NPs.}  
     \label{fig2:simulations}
\end{figure}

The unique ability of dielectric Mie voids to confine light in air (Fig.~\ref{fig2:simulations}~(a), inset of an empty void) is a major advantage for biosensing because it facilitates the access of analyte molecules to the regions of electromagnetic field enhancement and allows more efficient use of sample volumes~\cite{Arslan2025}. Contrary to most high-index dielectric resonators, voids require strong absorption in their material to form a pronounced optical Mie mode~\cite{Hentschel2023}. Therefore, in our work, we utilized Mie voids made of silicon, which provides large optical losses in the visible range\cite{Aspnes1983}. Furthermore, silicon wafers are low-cost and there are well-established fabrication protocols for patterning and etching. The voids were arranged in a square array of non-interacting resonators with a spacing of 200~nm (period $\Lambda = 700$~nm in both directions). The signal readout is defined by the interference of the large non-resonant reflection from the silicon substrate and the resonant scattering of light by the voids. The resulting reflection spectrum, therefore, features a dip corresponding to the Mie modes of the voids (Fig.~\ref{fig2:simulations}~(a), gray dashed line). 

We optimized the design so that at the resonance wavelength the structure suppresses the reflected signal, thus providing a homogeneously dark background, which is beneficial for high contrast imaging of bound nanoparticles on its surface. Our digital biosensing approach utilizes gold nanoparticles as contrast agents. The NPs bound to the void cause strong disturbances of the resonant optical mode, yielding a high contrast signal for each NP over the broadband low reflectivity of the nanostructured substrate. Furthermore, the use of gold nanoparticles grants access to the well-developed portfolio of functionalization and protein conjugation protocols\cite{jazayeri2016various, tam2017comparison}. 

The mechanism of NP detection is illustrated in Fig.~\ref{fig2:simulations}(a).  The efficiency of nanoparticle detection on the Mie void surface depends not only on the optimal spatial overlap of the NP with the optical mode field pattern, but also on the spectral overlap between the mode and the plasmonic resonance of the NP. We chose the standard commercially available gold nanoparticle size of 100~nm, which features a plasmonic resonance at $\lambda_{NP} = 525$~nm in air. When developing the void design, we targeted the largest change in scattering strength of the void-NP system at this wavelength as an indicator of their efficient interaction (see Methods for details). The optimization of the geometrical parameters of the void yielded values of diameter $d = 500$~nm and depth $h = 160$~nm. The reflectance spectrum from an array of these voids was modeled in COMSOL Multiphysics (see Methods for details) and is shown in Fig.~\ref{fig2:simulations}~(a) with a gray dashed line. As expected, this spectrum displays a prominent dip around the target wavelength $\lambda_{NP}$. The introduction of a NP leads to an increase in reflectance, the magnitude of which depends on the position of the nanoparticle inside the void (Fig.~\ref{fig2:simulations}~(a), yellow and orange curves). Notably, the maximum change in signal was achieved when the incident light polarization matched the direction of the NP shift from the center of the void. For this reason, in the experimental setup, we utilized unpolarized light to cover all possible displacements of a particle. 

Importantly, the field patterns shown in the insets of Fig.~\ref{fig2:simulations}(a) indicate that the Mie void mode also partially extends to the region above the spacing between the voids. This allows the detection of nanoparticles that are bound to the sensor surface outside of the voids. Our simulations show that in this case, the reflectance signal from the unit cell is even lower than that from an empty void (Fig.~\ref{fig2:simulations}~(a), violet curve). Therefore, our engineered design fully utilizes the entire sensors surface, in contrast to previously reported designs, where certain regions are insensitive and fail to detect some NPs ~\cite{Jing2019, Belushkin2018, Belushkin2020, VanDongen2022}. 

For the nanofabrication of the Mie void arrays based on the developed design, we used two fabrication approaches. During the verification of the detection principle, as well as for training the CNN algorithm with labeled images, we employed the e-beam lithography (EBL) technique. EBL offers extremely precise patterning and flexibility in design modifications (see Methods for details), but due to its low-throughput with sequential writing, EBL is inherently incompatible with low-cost and scalable fabrication routines. Therefore, to increase the transferability of the presented platform, we also developed fabrication protocols with deep-UV lithography for the patterning of the Mie voids (see Methods for details). A SEM image of the structures fabricated with DUV lithography is presented in Fig.~\ref{fig2:simulations}~(b) and shows comparable quality to EBL fabricated arrays, further confirmed by spectroscopic measurements.

\section{Optical imaging}

As discussed above, the simulation results demonstrate that the presence of plasmonic nanoparticles on the Mie voids surface induces a reflectance change (Fig.~\ref{fig2:simulations}~(a)). Therefore, optical imaging of the sensor surface in reflection mode should indicate the binding of the NPs by a characteristic local change in reflectivity, which eventually enables the digital detection of biomarkers. To showcase this detection principle, we deposited 100-nm gold particles onto the Mie voids array using simple incubation of the NP solution on the surface, which resulted in a homogeneous and random distribution of NPs located both inside and outside of the voids. This was confirmed using scanning electron microscopy (SEM, Fig.~\ref{fig2:simulations}b), which allowed us to precisely locate the actual positions of the nanoparticles that were used later for cross-checking the optical imaging data.

\begin{figure}[t!]
    \centering
    \includegraphics[width=0.99\columnwidth]{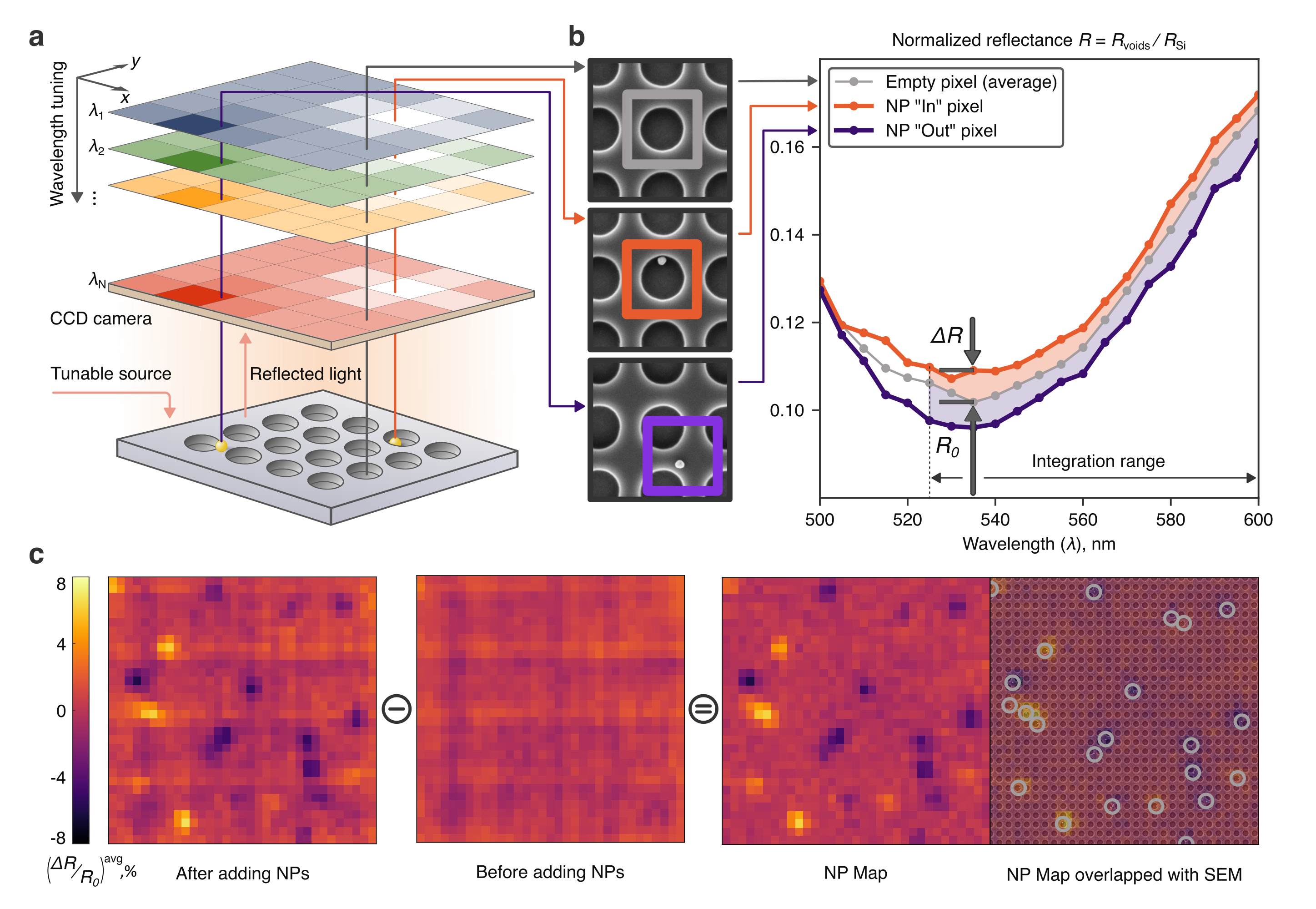}
    \caption{\textbf{Optical imaging.} a) Schematic view of the hyperspectral imaging configuration used to analyze the optical response of Mie voids with individual NPs. The Mie voids array is illuminated with a narrow-band tunable source. The reflected light is collected by a CCD camera matrix, forming a hyperspectral cube -- a stack of images, each corresponding to a certain illumination wavelength. Areas of the Mie void array with Au NPs projected onto the camera matrix generate either brighter (NP "In") or darker (NP "Out") pixels over a broad wavelength span. b) Experimental reflectance spectra extracted from the acquired hyperspectral cube. Each point in this graph takes the intensity value of a certain pixel (corresponding to the areas with NP either "In" or "Out") from a given wavelength $\lambda_i$ image. The reflectance $R_0$ from empty pixels (averaged) is shown with a thinner gray line. The presence of a nanoparticle induces reflectance change, $\Delta R$. The semitransparent region denotes the integration range used for condensing hyperspectral data into final NP maps. The insets show SEM images of the areas corresponding to the extracted pixels. The squares depict the size of a camera pixel relative to the magnified image of the voids. c) Average differential reflectivity $(\Delta R/R_0)^{\mathrm{avg}}$ maps. The NP map is obtained by subtracting the 'Before adding NPs' image from the 'After adding NPs' one. The final NP map is overlapped with a SEM image of the corresponding area to confirm the origin of bright and dark spots. Gray circles denote positions of NPs in the SEM image.}  
     \label{fig3:optical_detection}
\end{figure}

To realize the digital sensing approach with Mie voids, we implemented an optical imaging setup as illustrated in Fig.~\ref{fig3:optical_detection}(a). It was based on a standard upright microscopy system, with the full field of view of the objective projected onto the CCD matrix. To gain detailed optical insights into the operation principle, it was useful to extract the relevant signal change arising from NP binding to our sensor surface with both spatially and spectrally resolved data, because the most prominent modifications of reflectivity happen close to the void resonance (Fig.~\ref{fig2:simulations}a). For this purpose,  we utilized a 'hyperspectral' imaging configuration \cite{khan2018modern, yesilkoy2019ultrasensitive}, where the spectrally narrow light with tunable wavelength illuminates the sample. For each wavelength, a reflectivity image is recorded, which eventually yields a hyperspectral data cube -- a stack of images, each corresponding to a different incident wavelength (Fig.~\ref{fig3:optical_detection}~(a)). In our work, we realized this scheme using unpolarized light from a supercontinuum laser with a tunable band-pass filter (width of the band $\approx$2 nm, see Methods for details). 

We limited the range of hyperspectral data collection to 500-600~nm, which covers the largest change in reflectance signal for our design according to the simulations (Fig.~\ref{fig2:simulations}a). The wavelength tuning step was set at 5~nm, which is sufficient given the broad linewidth of the Mie void resonance. Figure~\ref{fig3:optical_detection}~(b) shows the reflectance spectra from single pixels extracted from the hyperspectral data. These pixels correspond to the nanoparticles located inside ("In") and outside ("Out") of the Mie void, which we confirmed by the SEM image (inset of Fig.~\ref{fig3:optical_detection}~(b)). Each point in these reflectance curves represents the intensity value for a given pixel in an image recorded at $\lambda_i$ excitation wavelength. The averaged spectrum of empty void pixels is shown for comparison with a gray line. In full agreement with the simulations (Fig.~\ref{fig2:simulations}a), it has an intermediate reflectivity signal between the "In" and "Out" spectra. 

In the post-processing routine, we condensed the acquired hyperspectral data into a single 2D map by calculating the average differential reflectivity $(\Delta R/R_0)^{\mathrm{avg}}$. In such a map, each pixel value at a location $(n, m)$ was calculated as follows:
\begin{equation}
    \left( \Delta R/R_0 \right)^{\mathrm{avg}}_{n,m} = \frac{1}{N}\sum_{i=1}^{N}\frac{R_{n,m}^i - R_0^i}{R_0^i}\cdot 100\%,
\end{equation}
where $R_0^i$ is the averaged background reflectance value for a given $\lambda_i$ image, $R_{n,m}^i$ is the value of the pixel with $(n,m)$ coordinates in a given $\lambda_i$ image, and $N$ is the total number of images taken for integration. From the SEM pictures, we observed that the vast majority of nanoparticles located inside the voids were positioned near the void edges. Therefore, we set the integration range to 525-600 nm, as it corresponds to the largest reflectance modulation for both "Out" and "In" edge NPs in the simulations (Fig.~\ref{fig2:simulations}~(a)). The final NP map was obtained by subtracting the differential reflectivity maps measured before and after the deposition of the nanoparticles (Fig.~\ref{fig3:optical_detection}~(c)). This subtraction allowed us to filter out signal variations caused by the inhomogeneity of the sample surface. The resulting NP map therefore features only the signals induced by the deposited nanoparticles -- bright spots (larger reflectance) corresponding to particles located inside the voids and dark spots (smaller reflectance) to the ones outside the voids. For confirmation of the actual position of the nanoparticles on the sample surface, we superimposed the NP map with the SEM image of the same area (Fig.~\ref{fig3:optical_detection}~(c)), which showed a very good correspondence between the signal variation in the optical map and the actual NP coordinates (marked with gray circles in the image).   

\section{CNN classification}
The resulting NP map already gives us a valuable insight into the approximate surface density of the nanoparticles. However, the realization of digital biosensing requires the calculation of the exact number of binding events, which is determined by the number of detected particles. Direct counting raises certain challenges to be overcome. Firstly, the image of a nanoparticle is blurred due to the finite resolution of the imaging setup. Secondly, the sampling of the sensor image by the pixels of the CCD camera (Fig.~\ref{fig3:optical_detection} (b) SEM insets, squares denote pixel size relative to the image of the Mie voids) leads to additional variations in the observed signal intensity change. The situation becomes more challenging at higher surface densities of the particles, which can form clusters manifested as signals of larger amplitude in the optical image and can cause spatial overlap between separate NP signals. Additionally, the measured maps exhibit small local deviations of the background reflectance from the mean value $R_0$. All these factors hinder the performance of standard image processing techniques, such as thresholding and filtering. To address these challenges, we applied a fully convolutional neural network (CNN), in particular, the U-Net architecture, which is a commonly utilized tool for image segmentation with a relatively small amount of training images\cite{ronneberger2015u}. 

\begin{figure}[t!]
    \centering
    \includegraphics[width=0.99\columnwidth]{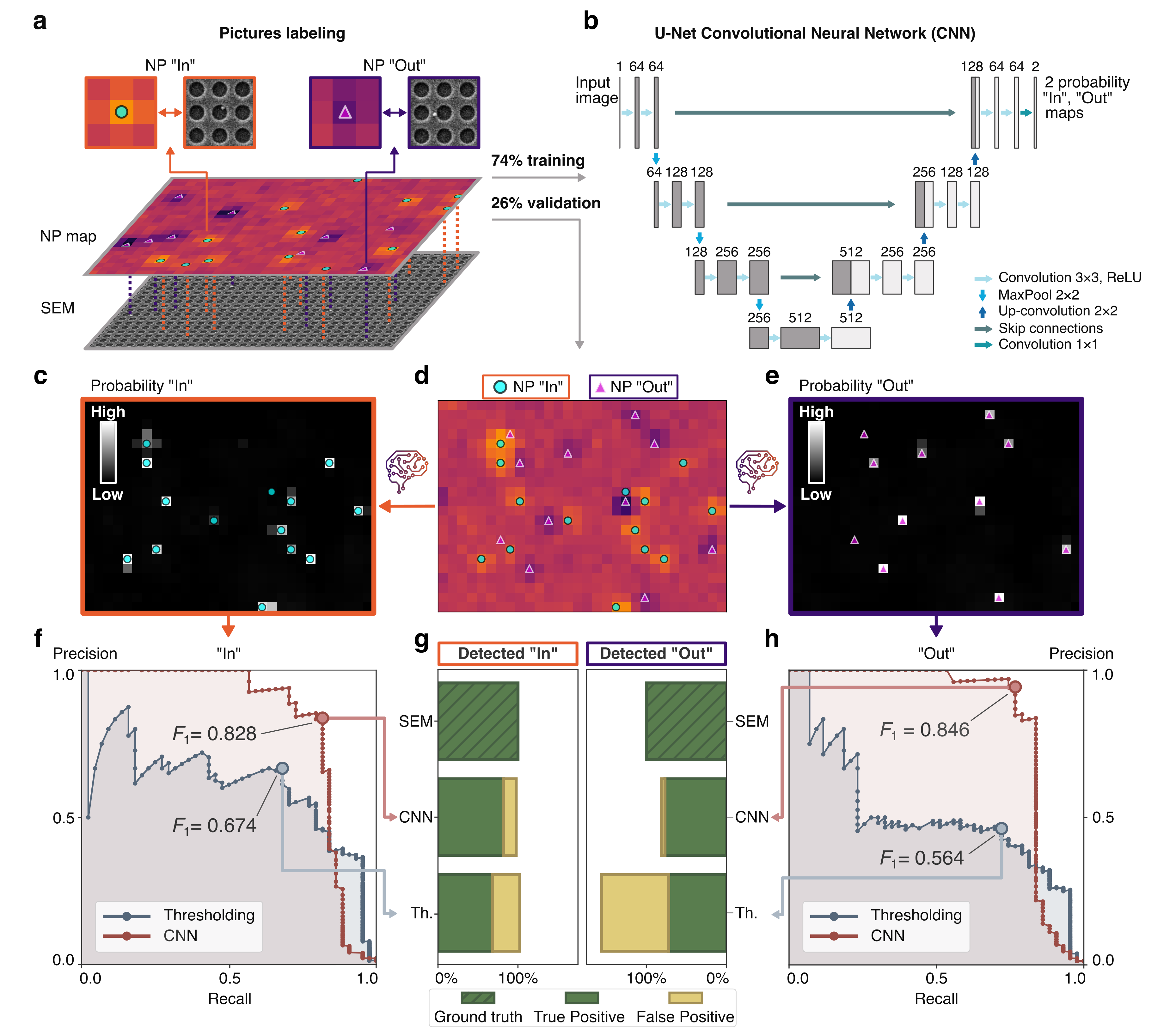}
    \caption{\textbf{CNN-Aided digital detection and counting of NPs on Mie voids.} a) Labeling of the NP maps using SEM images for training and validation of the model. The maps are marked with 'ground truth' pixel locations acquired from SEM images for "In" and "Out" NPs. b) Illustration of the U-Net architecture of the CNN. The framework utilizes three hidden layers. The numbers above each node indicate the number of applied convolutional filters at this step. The NP map from the validation dataset (d) is processed with the trained CNN model to generate two probability maps for "In" (c) and "Out" (e) NPs. 'Ground truth' locations of the respective NPs are indicated with cyan circles in the "In" image and purple triangles in the "Out" image. f, h) Comparison of precision-recall curves between the CNN-based algorithm applied to "In" (f) and "Out" (h) probability maps and simple thresholding of the original image. Markers on the curves indicate the highest $F_1$-score values for both methods. g) Combined statistics of the performance of both methods on the validation dataset.}  
     \label{fig4:CNN}
\end{figure}

The U-Net framework that we employed consisted of a contracting and an expansive path with three hidden layers (Fig.~\ref{fig4:CNN}~(b)). Each layer applied the following sequence to the input channels: - convolution with 3$\times$3 in-plane kernel; - application of the rectifying linear unit (ReLU) activation function; - repetition of the same cycle for a second time. The number of channels, which corresponds essentially to the number of applied convolutional filters, increased with the network depth. In the contracting path, the feature maps were downsampled between the layers via max pooling, whereas in the expansive path, the maps were upsampled to the higher-resolution levels. The input of each layer in the expansive path was formed by concatenating the upsampled feature maps with those passed through skip connections. This allowed the network to combine low- and high-level features at each stage. The network was designed to deliver two types of extracted features, corresponding to "In" and "Out" probability maps (see Methods for details). 

Prior to the training procedure, we performed labeling of the input NP maps, 24$\times$32 pixels each, by superimposing them with the corresponding SEM images and thus acquiring the 'ground truth' coordinates of $\sim$350 NPs (Fig.~\ref{fig4:CNN}~(a)). Approximately half of these nanoparticles were located inside the voids (marked with cyan circles in Fig.~\ref{fig3:optical_detection}) and the other half in between the voids (marked with purple triangles). Subsequently, we used 74\% of the labeled NP maps for training our CNN. To expand the training set synthetically, we added random data augmentation, including mirror flips and rotations of the images by up to 15 degrees. 

The evaluation of the CNN performance was completed with the validation image dataset (the remaining 26\% of images not yet seen by the network). For a single NP map input (Fig.~\ref{fig4:CNN}~(d)), the network generated two probability maps, corresponding to inside ("In", Fig.~\ref{fig4:CNN}~(c)) and outside ("Out", Fig.~\ref{fig4:CNN}~(e)) particles. One can see from these maps that the contrast with a background for 'ground truth' pixels was significantly enhanced compared to the original image. Therefore, it is possible to set a threshold and filter out the correct positions more accurately than by applying a thresholding procedure directly to the NP map. 

The final output of the model after thresholding can be referred to as a binary classifier since every single pixel of the input NP map is assigned either an 'NP' or 'Empty' label based on the set threshold. By comparing the 'ground truth' locations of the nanoparticles with the model output, we can distinguish the following categories of predicted values: true positive (TP) -- for a pixel with the correct 'NP' label assignment, false positive (FP) -- for a pixel wrongly assigned the 'NP' label and false negative (FN) -- for a pixel wrongly assigned the 'Empty' label. To quantify the classifier performance, we utilize the concept of precision and recall defined as follows:
\begin{equation}
    \mathrm{Precision} = \frac{\mathrm{TP}}{\mathrm{TP + FP}}; \quad \mathrm{Recall} = \frac{\mathrm{TP}}{\mathrm{TP + FN}}.
\end{equation}
Essentially, precision specifies the fraction of TP events among all the retrieved instances, whereas recall determines the fraction of the TP events among the 'ground truth' events. By changing the threshold value, one of the characteristics can be improved, while sacrificing the other.

A standard way to assess the performance of a binary classifier is to plot the precision-recall curve by varying the threshold values. The performance metric in this case is the area under the curve (AUC): the closer it is to 1, the better the classifier performs. In our case, we could set the thresholds for "In" and "Out" NP maps independently and, therefore, treated these two cases separately. We processed the NP maps from the validation dataset with our model (orange and violet arrows in Fig.~\ref{fig4:CNN} indicate the process flow for "In" and "Out" particles correspondingly) and compared the resulting precision-recall curves with those obtained by applying a thresholding procedure directly to the input NP maps (Fig.~\ref{fig4:CNN}~(f, h)). Notably, for both "In" and "Out" categories, the CNN drastically improves the detection performance, as the AUC of the CNN  (transparent brown areas) increases by 37\% and 65\%, compared with simple thresholding (transparent gray area) for inside and outside particles, respectively.

Depending on the application, any threshold can be chosen to gain either higher precision or recall. In our case, we aimed to balance both, and for this purpose, we calculated the $F_1$ score, which is the harmonic mean between the two:
\begin{equation}
    F_1 = \frac{2}{\mathrm{precision}^{-1} + \mathrm{recall}^{-1}} = 2\cdot\frac{\mathrm{precision}\cdot\mathrm{recall}}{\mathrm{precision} + \mathrm{recall}}.
\end{equation}
Accordingly, we established the thresholds that provide the highest $F_1$ score (highlighted with markers on the precision-recall curves in   Fig.~\ref{fig4:CNN}~(f, h)). Similarly to the AUC, the CNN improved the $F_1$ score for both "In" (by 22.8\%) and "Out" (by 50\%) particles, compared with simple thresholding. For the optimal threshold values, we performed digitization of the original NP maps from the validation dataset using both approaches. The data for "In"  and "Out" particles are summarized across the entire validation dataset in Fig.~\ref{fig4:CNN}~(g). As expected from its considerably higher $F_1$ metric, the CNN-based model produces generally more TP and fewer FP results, with the difference especially notable for the "Out" particles. Overall, our CNN model enabled us to detect $\sim$80\% of all the nanoparticles on the surface with as few as $\sim$10\% of false positive results.

\section{Biomarkers detection}

The introduced Mie voids platform, aided by artificial intelligence, opens broad opportunities for ultrasensitive biosensing applications. One of its key strengths lies in the visualization and counting of individual binding events, thus empowering the detection of target analytes that are present at very low concentrations in biological samples. To demonstrate the sensitivity of the developed platform, we performed detection of interleukin-6 (IL-6), which is a pro-inflammatory cytokine involved in immune regulation, infection response, and chronic inflammation.  Moreover, elevated concentrations of IL-6 have been associated with cardiovascular risks\cite{Mehta2024} and the early development of certain cancers\cite{Lippitz2016}. 
Therefore, IL-6 represents an important disease biomarker. However, the clinical concentrations of IL-6  are typically very low within a few tens to hundreds of pg/mL range\cite{Said2021, Macy1996},  which makes its accurate quantification particularly challenging. In this regard, the detection of IL-6 provides a compelling demonstration for the capabilities of our sensor platform.

\begin{figure}[h!]
    \centering
    \includegraphics[width=0.99\columnwidth]{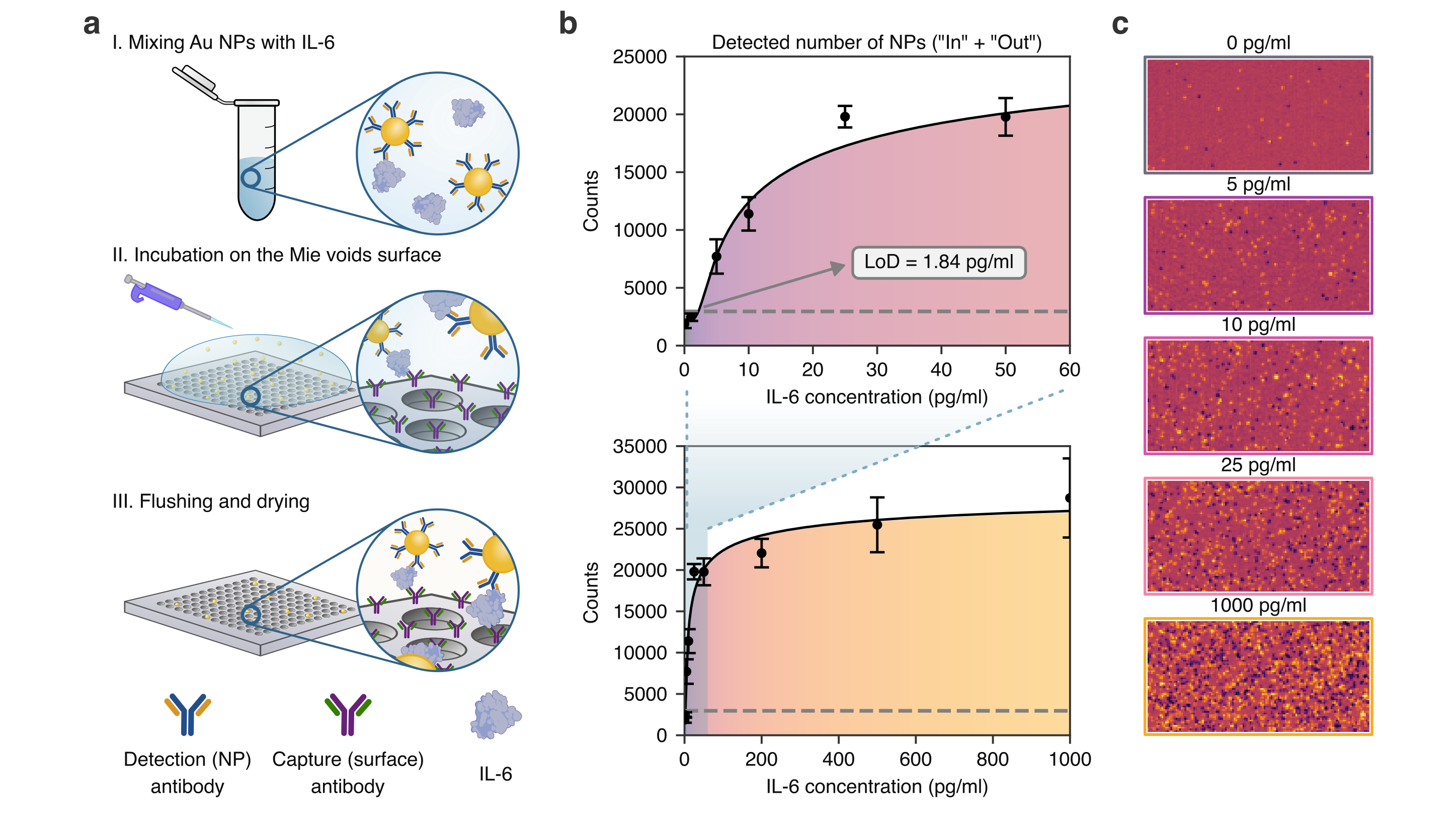}
    \caption{\textbf{IL-6 biomarker detection.} a) Sample preparation procedure. I. Pre-functionalized NPs are mixed with IL-6 molecules in a buffer. IL-6 molecules bind to NPs via antibody-analyte recognition. II. The resulting solution is deposited onto the Mie void chip and left for incubation for one hour. NPs with IL-6 molecules get immobilized on the sensor surface through capture antibody-analyte recognition. III. The chip is washed to remove unbound NPs and dried before measurement. b) Calibration curve showing the detected number of NPs (both "In" and "Out") versus the concentration of IL-6. The output signal of the digital biosensing platform is presented with the mean values at each concentration point, with the error bars showing  $\pm$s.d. (standard deviation). The limit of detection is defined as three times the standard deviation of the blank signal (0 pg/ml) counted from the mean value, and is depicted with a gray dashed line. c) Snippets from NP maps corresponding to different concentrations of IL-6. The number of immobilized NPs increases with increasing concentration.}  
     \label{fig5:IL-6}
\end{figure}

In the previous sections, we showed that the developed approach enables the counting of single nanoparticles on silicon Mie void arrays with high precision. To ensure that each detected particle can be associated with the presence of one or more IL-6 molecules, we utilized a sandwich immunoassay. In this assay, the analyte is 'sandwiched' between two antibodies, with one binding site attached to the detection antibody conjugated to the nanoparticle and the other site attached to the capture antibody conjugated to the chip surface (Fig.~\ref{fig5:IL-6}~(a)). In such a scheme, a gold nanoparticle can be immobilized on the Mie void array if at least one analyte (IL-6) molecule is bound to the detection (NP) antibody. The proposed detection mechanism was implemented using a three-step approach. First, a solution of nanoparticles functionalized with detection antibodies was mixed for one hour with IL-6 molecules in phosphate-buffered saline (PBS), to ensure the recognition of the analyte molecules (Fig.~\ref{fig5:IL-6}~(a), step I, see Methods for more details). Next, the silicon Mie void chips with capture (surface) antibodies were incubated for one hour in the resulting solution (Fig.~\ref{fig5:IL-6}~(a), step II, see Methods for more details). This was followed by washing the chip with Milli-Q water to remove unbound nanoparticles and drying it with $\mathrm{N}_2$ stream (Fig.~\ref{fig5:IL-6}~(a), step III, see Methods for more details) for subsequent optical measurements. As a result, most of the particles remaining on the chip were bound through antibody–analyte interactions (specific capture), in contrast to nonspecific adhesion, where nanoparticles attach to the surface without molecular recognition. In Fig.~\ref{fig5:IL-6}~(c), examples of optical NP maps are shown, derived at different concentrations. Notably, the "blank" sample measurements (0~pg/ml, no IL-6 added in the prepared samples, top row) demonstrated only a small number of detected nanoparticles (bright and dark spots), which clearly increased with higher concentration of the analyte due to specific capture events.  

The dependence of the detected number of particles on the IL-6 concentration is represented by the calibration curve (Fig.~\ref{fig5:IL-6}~(b)). Each point on this curve corresponds to a specific IL-6 level and was obtained by applying the trained CNN model to the optical NP maps. For each entry, we post-processed areas of 800$\times$800 pixels to detect both "In" and "Out" bound NPs. Each measurement was repeated several times using independent depositions to evaluate the average output value and its dispersion (represented by the error bars). We assessed the nonspecific signal through "blank" sample measurements (0~pg/ml) with the same density of the nanoparticles without adding the IL-6 molecules. The limit of detection (LoD) of the platform was defined as the mean "blank" value plus three standard deviations. Importantly, our AI-assisted digital biosensing approach provided an LoD value as low as 1.84~pg/ml, which falls within the physiological IL-6 range for healthy individuals\cite{Said2021}. To account for a potential baseline drift in real-world applications, we incorporated two distinct areas into the chip design for reference and sensing measurements (Fig.~\ref{fig1:intro}~(b)), one serving as a negative control and the other as the detection area. Here, the compatibility of the design with DUV lithography becomes a critical feature, as it allows for low-cost and scalable fabrication of sensor chips.

\section{Conclusions}

We have presented a novel nanophotonic platform based on silicon structures, that combines contrast-enhanced detection with a deep learning approach for ultrasensitive digital biosensing. In our structures, light is resonantly confined in the Mie voids while also extending to the surface in between them, providing an efficient interaction with captured nanoparticles across the entire sensor area. This interaction creates local perturbations in the reflected signal, manifesting as high-contrast signatures of individual binding events in the optical image of the biosensor and indicating the presence of the analyte molecules. The detection performance is boosted by applying the CNN framework to the acquired optical images yielding a detection rate of around 80\% for the particles on the sensor surface with as low as 10\% false positive retrievals. 

The use of gold nanoparticles as contrast agents enables the measurement of biomarkers digitally, which improves the limit of detection compared to conventional ensemble-averaged refractometry or intensity-based sensing methods. Our proof-of-concept demonstration has showed the detection of an important protein biomarker at concentrations as low as 1.84 pg/ml from very small sample volumes ($\sim5\mu$L). The adaptation of the DUV lithography technique significantly simplifies the transferability towards mass-scale fabrication and increases its potential for translation to real-world applications. 

We notice that alternative designs and shapes of dielectric Mie voids might be explored for different excitation wavelengths. The simplicity of the optical readout in the presented study provides prospects for developing compact point-of-care devices for rapid and reliable diagnostics. Particularly, we envision that the bulky hyperspectral setup can be substituted with either a narrow-band LED or a broadband illumination combined with a spectral filter to match the integration window. In this paper, we have showcased biomarker detection with the medically relevant inflammation protein IL-6, but we envision that the introduced sensing platform can be extended to detect other target analytes by switching the capture/detection probes and surface functionalization protocols, while keeping the optical framework unchanged. We believe that our novel platform for digital optical biosensing, empowered by dielectric Mie voids and enhanced with a convolutional neural network, creates a versatile tool that opens up new possibilities for a broad range of bioanalytical applications.  

\section{Methods}
\subsection{Numerical simulations}
Numerical simulations were performed with COMSOL Multiphysics\textsuperscript{\textregistered} software. To optimize geometric parameters of a Mie void, we utilized a two-dimensional axisymmetric electromagnetic model mimicking a full three-dimensional system\cite{Gladyshev2024}. The physical domains included a silicon substrate (optical functions for silicon were taken from Ref.~[\citen{Aspnes1983}]) with a cut-out void made of air, both appearing as rectangles in two-dimensional projections, as well as top air medium, simulated as a quarter of a circle.  A spherical gold nanoparticle (optical functions for gold were taken from Ref.~[\citen{McPeak2015}]), appearing as a semicircle in a two-dimensional model, was located strictly in the center of a void to preserve the symmetry, and placed 7 nm above the bottom surface to simulate for functionalization layer and avoid mesh inconsistencies near the contact point. The physical domains were surrounded by perfectly matched layers to absorb incoming radiation and imitate infinite space. The whole system was illuminated with a normally incident plane wave, accounting for transmission and reflection from a plain silicon substrate. For normal incidence, the scattering properties were rigorously modeled using a single azimuthal order $m = 1$. Scattered light was collected in an upper arc to evaluate the half-space scattering cross-section. 

The simulation of the periodic array of voids was performed in a full three-dimensional geometry utilizing Floquet boundary conditions. The air domain and silicon substrate in a unit cell were simulated as rectangular blocks and backed with perfectly matched layers to absorb incoming radiation and avoid re-reflections. A cylindrical Mie void of air was cut out from the silicon substrate domain. The incident radiation was modeled with an excitation periodic port, which also collected reflected intensity. 

\subsection{Nanofabrication}

{\bf Electron-beam lithography}. The samples used for optimization of image processing algorithms were fabricated by means of e-beam lithography. We spincoated electron beam resist - PMMA (polymethyl methacrylate) with 950.000 molecular weight - on the preliminarily cleaned silicon chip surface and baked it for 5 minutes at 180 \textsuperscript{o}C yielding the total resist thickness of 180~nm. The resulting chip was patterned with electron beam, followed by developing the resist in methyl isobutyl ketone (MIBK), diluted in IPA at 1:3 ratio, to remove exposed parts. Further step included the transfer of the design to the silicon substrate using fluorine-based deep reactive ion (DRI) etching. Time of etching was optimized to obtain the required depth. To get the final structures, we stripped the resist mask using Remover 1165 and $\mathrm{O_{2}}$-plasma.

{\bf Deep UV lithography}. The chips for final biosensing experiments were prepared using wafer-scale deep-UV lithography technique. Preliminarily cleaned standard 4-inch silicon wafer was automatically processed in TEL Clean Track ACT-8 coater and developer -  first layer of HMDS was deposited to increase adhesion of the next layer of positive tone deep UV resist (M108Y), resulting in the total thickness of 400~nm. The resist was exposed in an in-line ASML PAS 5500/350C stepper with UV light from KrF laser source (248 nm). The light came through the reticle photomask and was focused with the lens at the wafer surface. The wafer was further post-exposure baked at 130 $^\mathrm{o}$C for 90 seconds and automatically developed for 60 seconds in TMAH to remove the exposed parts. Consequent step repeated the procedure of transferring the design to the silicon wafer with DRI etching, same way as in e-beam based technique. The resist layer was removed with subsequent SVC-14 stripper and high power (500~W) $\mathrm{O_{2}}$-plasma. For the convenience of biosensing experiments, the full wafer, preliminarily coated with 3~$\mu$m thick protective photoresist layer, was diced into 24 square chips 1.5 cm$\times$1.5 cm each. Protective layer from the resulting chips was stripped using SVC-14 and $\mathrm{O_{2}}$-plasma.

\subsection{Hyperspectral imaging}
Optical measurements of reflectance from the Mie voids sensor were performed with hyperspectral imaging setup. Illumination scheme involved a supercontinuum laser (NKT Photonics, SuperK EXTREME EXR-15) coupled to a tunable bandpass filter (NKT Photonics, LLTF), which was brought into an upright microscope (Nikon, Eclipse-Ti series) excitation channel. Collimated unpolarized laser line (bandwith $<$2.5~nm) hit the sample and reflected light was collected with the same 20$\times$, NA~0.4 objective from the top. The 1024$\times$1024 pixels images of the sample were recorded with EMCCD camera (ANDOR, iXon Ultra 888) with acquisition time of 120 ms and gain of 10. By sweeping the incident wavelength over the spectral range of interest, the hyperspectral cube was obtained, which was further normalized to the data cube corresponding to a bare silicon substrate. The devices were controlled using a centralized Matlab code.

\subsection{Image post-processing}
Differential reflectivity maps were obtained using the NumPy package in Python. For each of the normalized images, corresponding to the excitation wavelength $\lambda_i$ within the integration range, we calculated the average reflectance value $R_0^{i}$ over a certain spatial area. Next, this value was subtracted from the whole image and resulting array was divided by $R_0^i$. Obtained images were summed up yielding a single map, further divided by the total number of images within the integration range.

For creating the final NP map, we needed to subtract the differential reflectivity maps from after and before nanoparticle deposition. Since the sample was taken in and out of the setup between the two measurements, the resulting images had to be rotated to align with each other. For this purpose we applied rotation matrix $R$ to one of the images with defined center of rotation $\vec{c}$ and rotation angle $\theta$. This operation was performed in Python using OpenCV and SciPy libraries. To define $\vec{c}$ and $\theta$, two marker points - voids of a smaller size providing larger reflectance - were chosen in each image with position vectors $\vec{u}_{1,2}$ and $\vec{v}_{1,2}$ correspondingly. Resulting image vectors then were $\vec{u} = \vec{u}_{2} - \vec{u}_{1}$ and $\vec{v} = \vec{v}_{2} - \vec{v}_{1}$ and corresponding rotation angle was calculated as $$\theta = \operatorname{arg}(u) - \operatorname{arg}(v).$$ Center of rotation is derived from the condition $\vec{c} - \vec{v}_1 = R(\theta)(\vec{c} - \vec{u_1})$ and therefore results in $$\vec{c} = \left[R(\theta) - I\right]^{-1}(R(\theta)u_1 - v_1),$$
where $I$ is identity 2$\times$2 matrix.

\subsection{Convolutional neural network} 
The U-net architecture of a convolutional neural network was realized with the PyTorch library in Python. Both contracting (encoder) and expanding (decoder) paths utilized double convolution sequence at each layer: two-dimensional convolution was applied $L_{\mathrm{out}}$ times to the input channels stack $M\times N \times L_{in}$, with kernel sizes of 3$\times$3$\times L_{\mathrm{in}}$; resulting stack of size $M\times N \times L_{\mathrm{out}}$ passed through nonlinear activation function $\operatorname{ReLU}(x) = \operatorname{max\{0,}x\}$; two-dimensional convolution was applied again with kernels of 3$\times$3$\times L_{\mathrm{out}}$ size followed by activation function $\operatorname{ReLU}(x)$. At the first layer in the contracting path, the number of input channels is $L_{\mathrm{in}} = 1$, which is simply an original NP map of $M\times N$ size with $M = 24$, $N = 32$. The number of output channels at the first layer is $L_{\mathrm{out}} = 64$, which is doubled with every next layer of the network reaching $L_{\mathrm{out}} = 512$ at the deepest level. Before being transferred to a subsequent layer, the output stack is downsampled using $2\times 2$ MaxPool operation, yielding in-plane dimensionality of $\tilde{M}\times\tilde{N}$ with $\tilde{M} = M/2$; $\tilde{N} = N/2$. 

Input of the layers in expanding path was formed by: 1) reducing the number of channels from the previous layer twice, using $1\times1$ in-plane kernel; 2) upsampling the images to increase in-plane dimensionality twice by duplicating the values in $2\times 2$ window;  3) concatenating this stack with the data from the same layer in contracting path. The number of output channels was reduced twice with each subsequent layer. The final operation included $1\times1$ in-plane convolution with $L_{\mathrm{out}} = 2$ to form two output probability maps.

For training and validation of the model, we did labeling of 19 images, $24\times32$ pixels each. These images were taken as subsets of the larger field of view and corresponded to the areas analyzed with scanning electron microscopy. We applied the Laplacian of Gaussian (LoG) method to the SEM images of these areas to define locations of majority of the NPs within the frame, while the missing ones were added manually. 'Ground truth' coordinates in the optical images were further recorded using the Python script, which defined the position of maximum absolute value pixel in the vicinity of the mouse button click. Labeling was performed separately for "In" and "Out" NPs, resulting in a total of 188 "In" and 163 "Out" recorded entries.  

Training of the model was performed with the images from a training dataset, comprising 14 images, which were chosen from a full set randomly. To synthetically expand a training set, the input image at each epoch was assigned a random rotation with the angle up to 15\textsuperscript{o} and a random (yes/no) mirror flip. Training procedure involved adaptation of the kernel weights using errors backpropagation with PyTorch in-built Adam optimization method. To prevent overfitting, loss function was calculated in parallel for both training and validation dataset images, the latter not yet seen by the model. We chose the total number of 5000 training epochs, after which the loss function for the validation dataset did not decrease anymore.

\subsection{Biosensing experiment}

{\bf Nanoparticles preparation}. For the realization of a sandwich immunoassay, the gold nanoparticles should have been functionalized with detection antibodies. We used NHS-activated 100 nm Au NPs conjugation kit from Cytodiagnostics. 60 $\mu$l reaction buffer was mixed with 48 $\mu$l solution (500 $\mu$g/ml) of IL-6 antibodies from Hytest (L395), diluted in phosphate-buffered saline (PBS) . 90 $\mu$l of resulting mixture was added to the NPs and shaken for 2 hours. Afterwards, we mixed 10 $\mu$l of quencher buffer with NP solution. Next step involved 30 minutes centrifuging (400~rcf) of NPs with 900$\mu$l of 1\% bovine serum albumin (BSA), serving as a blocking agent, diluted in buffer (0.05\% Tween20 in PBS). This was followed by discarding the supernatant containing unbound proteins. Centrifuging and discarding steps were repeated for 5 times to ensure proper filtering of the unbound proteins from the solution. The total of 100~$\mu$l of final solution of functionalized NPs had optical density $\mathrm{OD}\approx20$. 

In analyte detection experiments, we prepared two stock concentrations, specifically 1.25 ng/ml and 40~pg/ml, of IL-6 in PBS buffer. The functionalized nanoparticles were mixed with IL-6 solution and PBS based on a desired concentration of analyte and the required density of the particles. After optimization steps, we chose to fix the density of the particles at OD~=~0.5. NPs with IL-6 were shaken for 1 hour to guarantee homogeneous mixing and binding of analyte molecules with functionalized NPs.

{\bf Mie voids functionalization}. For functionalization of native silicon oxide layer on the chip surface we utilized silanization technique. The monolayer of (3-glycidoxypropyl) trimethoxysilane (3-GPS, from Sigma-Aldrich) was used to immobilize capture antibodies. The silane end of the 3-GPS molecule binds to the silicon oxide surface, whereas its functional group attaches to the amine group of the antibody. 

The chips were cleaned in RCA solution for 30 minutes at 50~\textsuperscript{o}C and dried at 80~\textsuperscript{o}C for 5 minutes. Prepared this way, the chips were shaken (3 minutes) and incubated (20 minutes) in 3-GPS solution in toluene (1\% volume concentration) in a glass chipholder, which was followed by shaking in fresh toluene twice for 5 minutes to remove unreacted molecules. The chips were further baked for 30 minutes at 120~\textsuperscript{o}C followed by final shaking in fresh toluene for 15 minutes and $\mathrm{N}_2$ drying. Immediately after chips preparation, we dropcasted capture IL-6 antibodies (HyTest, L519, 1~mg/ml in PBS buffer) and left the surface for overnight incubation at 4~\textsuperscript{o}C in a humid environment (closed Petri dish with the water droplets) to avoid evaporation. Next day, the chips were rinsed with Milli-Q water, $\mathrm{N}_2$ dried, and incubated in BSA solution (10~mg/ml, in a humid environment) for 1 hour at room temperature to block unoccupied sites and prevent non-specific binding. Subsequently the chips were rinsed with Milli-Q water and $\mathrm{N}_2$ dried again to prepare for IL-6 biosensing experiment.

{\bf IL-6 detection experiments}. To confine the incubation solution within the sensing areas, we attached silicone rings (Sigma-Aldrich) on top of the prepared chips. The solution of IL-6 conjugated with functionalized Au NPs at OD = 0.5 was taken immediately after a mixing step and dropcasted in the silicone wells (5~$\mu$l/well). The chips were incubated for 1 hour at room temperature in a humid environment (closed Petri dish with the water droplets) to let the NPs settle down and specifically attach to the surface via capture antibody-analyte recognition. Finally, the sensors were thoroughly rinsed with Milli-Q water to wash away the unbound particles and  $\mathrm{N}_2$ dried before the optical measurements. 

\section{Acknowledgments}
The authors are thankful to the CMi staff who assisted with fabrication process flows development and instruments handling.
D.R. acknowledges Ruslan Gladkov for his assistance with CNN algorithm development; Albert Dominguez for fruitful discussions on image processing tasks; and the team of the 'Image Processing' course in EPFL, specifically Micha\"el Unser and Dimitri Van De Ville, for introducing fundamentals and applications of image processing methods, some of which were utilized in this work.

D.R., A.S., W.Y., I.S. and H.A. acknowledge the use of nanofabrication facilities at the Center of MicroNano Technology of École Polytechnique Fédérale de Lausanne and acknowledge financial support from École Polytechnique Federale de Lausanne, the European Union’s Horizon 2020 research and innovation programme under the Marie Skłodowska-Curie grant agreement No. 955623 (H2020MSCA-CONSENSE), the Swiss National Science Foundation (SNSF, Sinergia grant CRSII5\_213534), and the Swiss State Secretariat for Education, Research and Innovation (SERI)  under the contract numbers 22.00018 and 22.00081 in connection with the projects from the European Union’s Horizon Europe Research and Innovation Programme under agreements 101046424 (TwistedNano) and 101070700 (MIRAQLS); and financial support of Innosuisse (grant no. 121.316 IP-LS). Y.K. was supported by the Australian Research Council (Grant No. DP210101292) and International Technology Center Indo-Pacific (ITC IPAC) via Army Research Office (Contract FA520923C0023). 

\bibliography{Mie_voids.bib}

\end{document}